\documentclass{elsartL}
\usepackage{graphicx}
\usepackage{amssymb}
\usepackage{amsmath}
\usepackage{mathrsfs}
\usepackage{physics}
\usepackage{mathtools}
\newcommand{\fsl}[1]{\ensuremath{\mathrlap{\!\not{\phantom{#1}}}#1}}
\begin{document}

\begin{frontmatter}
\title{Interactions relevant to the decoupling of the neutrini/antineutrini in the early Universe}
\author{E.~Matsinos}

\begin{abstract}
The interactions of the neutrini and antineutrini amongst themselves, as well as the interactions of these particles with the electrons and the positrons, are of interest in simulations of the early Universe and in studies of 
the processes involving compact stars. The objective in this paper is to create a reliable source of information regarding the differential and total cross sections of these interactions; expressions for these observables will 
be obtained using standard methodology. A number of relevant discrepancies in the literature will be addressed.\\
\noindent {\it PACS:} 13.15.+g, 14.60.Lm, 14.60.St, 98.80.Cq

\end{abstract}
\begin{keyword} neutrino, differential cross section, total cross section, decoupling, early Universe
\end{keyword}
E-mail: evangelos[dot]matsinos[at]sunrise[dot]ch
\end{frontmatter}

\section{\label{sec:Introduction}Introduction}

Despite the fact that the expressions for the cross sections, corresponding to the interactions of the neutrini and antineutrini (henceforth, (anti)neutrini for short) with the ingredients of the plasma of the early Universe, 
are straightforward to obtain, the information retrieved from several sources in the literature, be they books, scientific publications, or presentations in Conferences, is frequently incorrect. To the best of my knowledge, 
there is no place in the literature where the correct formulae, pertaining to these scattering processes, are listed in a manner which is not subject to misinterpretation and misunderstanding.

Such interactions are interesting not only in the context of simulations of the early Universe, but also in terms of the physical processes taking place in the collapsing cores of supernovae. In fact, the early papers on this 
subject had been stimulated by studies of the physics of supernova explosions. The neutrino-antineutrino annihilation to electron-positron pairs has been investigated for one additional reason, namely in terms of the possibility 
that it provides a driving mechanism for the creation of gamma-ray bursts in compact stars, i.e., in white dwarves, neutron stars, and black holes.

The goal in this paper is to create a reliable source of information with respect to these interactions, as well as to discuss a number of relevant discrepancies in the literature. One additional reason for writing this paper 
is that only convenient high-energy approximations are quoted in most cases for the interactions of the (anti)neutrini with the electrons and the positrons of the plasma, i.e., results obtained from calculations ignoring the rest 
mass of these particles; both the exact formulae and their approximated expressions at high energy are given in this work.

The structure of this paper is as follows. Sections \ref{sec:Conventions}-\ref{sec:Preliminary} provide the tools required for the evaluation of the cross sections. The cross sections are derived for all relevant processes 
in Section \ref{sec:Derivation} and are presented in tabular and schematic forms in Section \ref{sec:Summary}. Section \ref{sec:Discrepancies} provides a discussion of those of the discrepancies in the literature which I am 
aware of. Section \ref{sec:Conclusions} briefly summarises the findings of this paper.

\section{\label{sec:Conventions}Notation and conventions}

In the present work, all physical variables and observables refer to the `centre of momentum' (CM) frame of reference (unless, of course, they are Lorentz-invariant quantities, in which case there is no need to specify a frame 
of reference). Only processes with two particles in both the initial and final states are considered (i.e., $2 \to 2$ scattering). Used are the following notations and conventions.
\begin{itemize}
\item The speed of light in vacuum $c$ is equal to $1$.
\item Einstein's summation convention is used.
\item $I_n$ denotes the $n \times n$ identity matrix.
\item $A^\dagger$ denotes the Hermitian (conjugate transpose) of a square matrix $A$.
\item $g_{\mu \nu}$ denotes the Minkowski metric with signature `$+ \, - \, - \, -$'.
\item $\gamma_{\mu}$ are the standard Dirac $4 \times 4$ matrices, satisfying the relation $\{ \gamma_{\mu}, \gamma_{\nu} \} = 2 g_{\mu \nu} I_4$ for $\mu, \nu \in \{ 0,1,2,3 \}$. In addition, $\gamma_0^\dagger = \gamma_0$ and 
$\gamma_\mu^\dagger = \gamma_0 \gamma_\mu \gamma_0 = -\gamma_\mu$ for $\mu \in \{1,2,3\}$.
\item The matrix $\gamma_5 \coloneqq i \gamma_0 \gamma_1 \gamma_2 \gamma_3$ enters the two projection operators to fermion states of definite chirality, i.e., the operators projecting a Dirac field onto its left- and right-handed 
components. The matrix $\gamma_5$ satisfies the relations: $\gamma_5^\dagger=\gamma_5$, $\gamma_5^2 = I_4$, and $\{ \gamma_{\mu}, \gamma_5 \} = 0$ for $\mu \in \{0,1,2,3\}$.
\item The quantity $\epsilon_{i_0 i_1 \dots i_n}$ is the Levi-Civita symbol, defined as follows:
\begin{equation*}
\epsilon_{i_0 i_1 \dots i_n} \coloneqq \left\{
\begin{array}{rl}
+1 & \text{if $(i_0, i_1, \dots, i_n)$ is an even permutation of $(0, 1, \dots, n)$}\\
-1 & \text{if $(i_0, i_1, \dots, i_n)$ is an odd permutation of $(0, 1, \dots, n)$}\\
0 & \text{otherwise}
\end{array} \right.
\end{equation*}
\item $s$, $u$, and $t$ are the standard Mandelstam variables. The quantity $s$ is equal to the square of the CM energy; $t$ is equal to the square of the $4$-momentum transfer; $u$ is equal to the square of the $4$-momentum 
transfer for interchanged final-state particles.
\item For a $4$-vector $a$, $\fsl{a} \coloneqq \gamma_{\mu} a^{\mu}$.
\item $k$ and $p$ are the $4$-momenta of the incident particles (projectile and target, respectively).
\item $k^\prime$ and $p^\prime$ are the $4$-momenta of the scattered particles.
\item $q \coloneqq k-k^\prime=p^\prime-p$ is the $4$-momentum transfer.
\item $\theta$ and $\Omega$ denote the scattering angle and the solid angle, respectively.
\item $m_e$ stands for the electron (and positron) rest mass.
\end{itemize}

\section{\label{sec:FD}Feynman graphs relevant to the decoupling of the (anti)neutrini}

It is generally believed that, between the temperatures of one hundred billion K ($T_1=10^{11}$ K), corresponding to a mere one-hundredth of a second after the Big Bang, and about $T_2=2 \cdot 10^{10}$ K, corresponding to the 
cosmological time of about $250$ ms, when the (anti)neutrini decoupled from the other ingredients of the primordial plasma~\footnote{So long as their collision rates exceeded the expansion rate of the Universe (identified as 
the Hubble parameter, which is naturally dependent on the cosmological time), the (anti)neutrini remained in thermal equilibrium with the ingredients of the plasma. At the time when their collision rates dropped below the 
expansion rate of the Universe, the (anti)neutrini escaped the plasma. The detachment of the (anti)neutrini from their former engagement as an active part of the plasma is known as `decoupling'.}, our Universe was a superdense 
mixture of neutrini, antineutrini, electrons, positrons, and photons. In contrast to the huge densities of these particles, a few baryons (protons and neutrons) were also present, about six such particles for every ten billion 
of photons. Of interest in the context of this work are the interactions of the (anti)neutrini with other (anti)neutrini, as well as with the electrons and the positrons, during the temporal interval corresponding to the 
temperatures $T_1$ and $T_2$.

\begin{figure}
\begin{center}
\includegraphics[width=15cm]{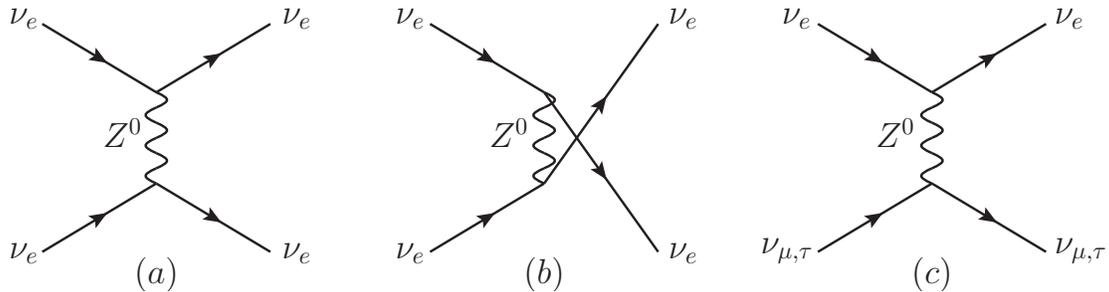}
\caption{\label{fig:NuNu}The permissible tree-level Feynman graphs for the interaction of an electron neutrino (projectile) with the neutrini of the electron, muon, and $\tau$-lepton generations of matter (targets). Graphs (a) 
and (b) relate to the scattering off the electron neutrino. Graph (c) relates to the scattering processes $\nu_e \nu_{\mu,\tau} \to \nu_e \nu_{\mu,\tau}$. In all cases, these $t$-channel exchanges involve the weak neutral 
current. Similar graphs are applicable in case of the muon and $\tau$-lepton neutrini as projectiles.}
\vspace{0.35cm}
\end{center}
\end{figure}

\begin{figure}
\begin{center}
\includegraphics[width=15cm]{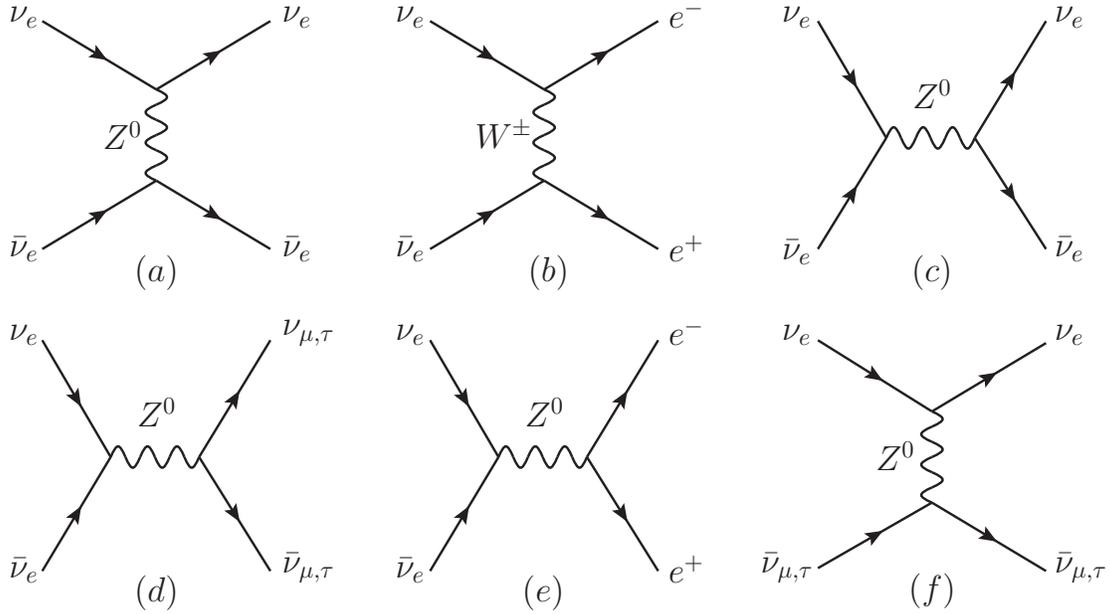}
\caption{\label{fig:NuNuBar}The permissible tree-level Feynman graphs for the elastic scattering of an electron neutrino (projectile) off the antineutrini of the electron, muon, and $\tau$-lepton generations of matter (targets). 
Graphs (a) and (b) relate to the $t$-channel scattering off the electron antineutrino, graph (a) via the exchange of the weak neutral current, graph (b) via the exchange of the weak charged current. Graphs (c)-(e) relate to 
the $s$-channel scattering off the electron antineutrino (annihilation graphs); in all such cases, only the weak neutral current is involved. Graph (f) relates to the $t$-channel scattering processes 
$\nu_e \bar{\nu}_{\mu,\tau} \to \nu_e \bar{\nu}_{\mu,\tau}$, where only the weak neutral current may be exchanged. Similar graphs are applicable in case of the muon and $\tau$-lepton neutrini as projectiles, save for graph 
(b) which cannot contribute in the energy range explored in this paper.}
\vspace{0.35cm}
\end{center}
\end{figure}

\begin{figure}
\begin{center}
\includegraphics[width=15cm]{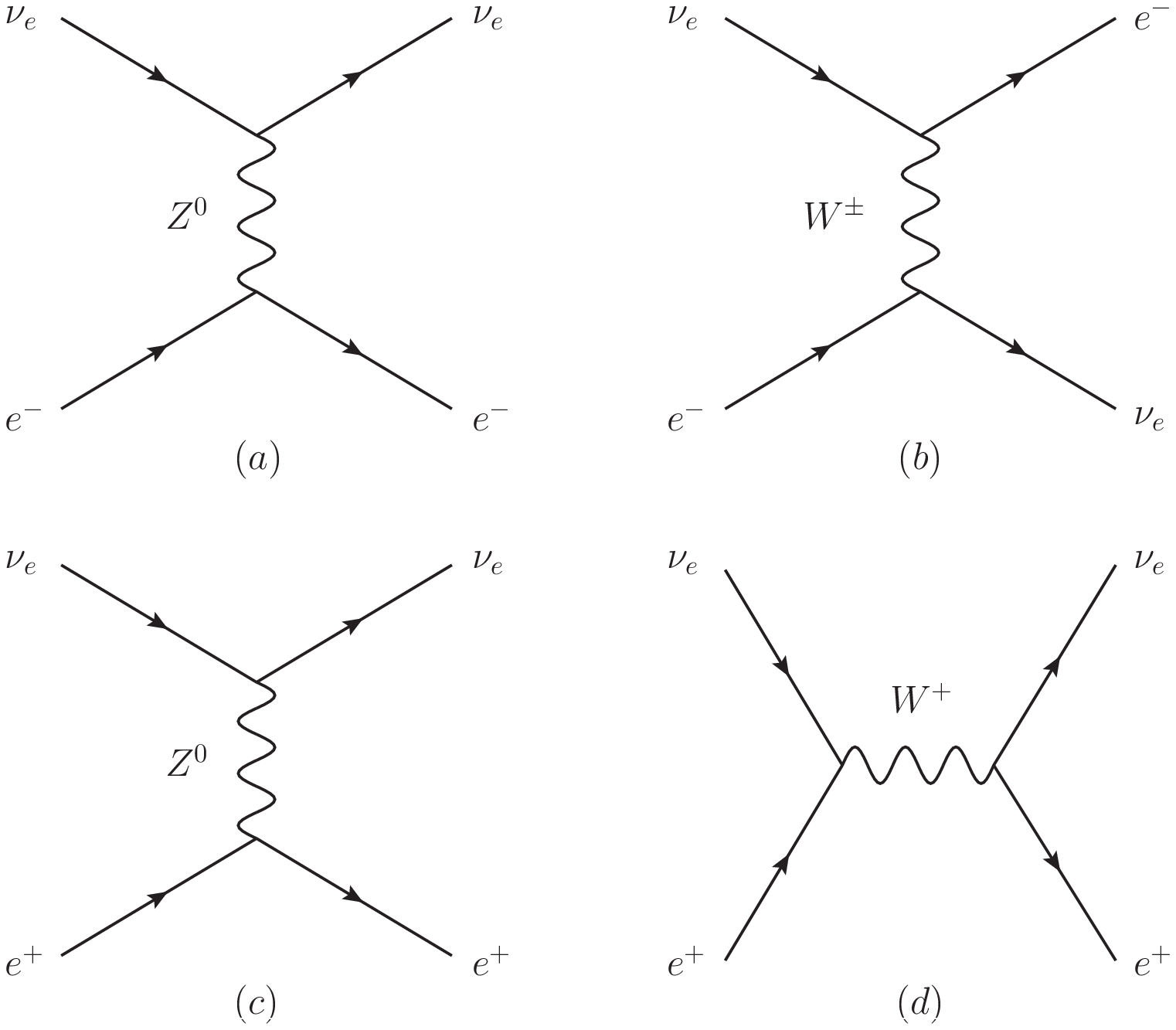}
\caption{\label{fig:NuElPos}The permissible tree-level Feynman graphs for the elastic scattering of an electron neutrino (projectile) off an electron and a positron (targets). Graphs (a) and (b) relate to the scattering off 
an electron, graph (a) via the $t$-channel exchange of the weak neutral current, graph (b) via the $u$-channel exchange of the weak charged current. Graphs (c) and (d) relate to the scattering off a positron, graph (c) via 
the $t$-channel exchange of the weak neutral current, graph (d) via the $s$-channel creation of a $W^+$ boson. In the case of the muon and $\tau$-lepton neutrini as projectiles, only graphs (a) and (c) contribute: graph (b) 
cannot contribute in the energy range explored in this paper, whereas graph (d) does not contribute on the basis of first principles (conservation of the leptonic number within each generation of matter).}
\vspace{0.35cm}
\end{center}
\end{figure}

The permissible tree-level Feynman graphs (henceforth, simply graphs) for the interaction of an electron neutrino (projectile) with the leptons of the plasma (targets), resulting in no more than an electron-positron ($e^+ e^-$) 
pair in the final state, are shown in Figs.~\ref{fig:NuNu}-\ref{fig:NuElPos}. All these graphs involve the exchange of either of the two intermediate vector bosons (IVBs) which are associated with the weak interaction, namely 
the $Z^0$ boson, which is associated with the neutral current (NC), and the $W^\pm$ boson, which is associated with the charged current (CC). Given that it yields a cross section several orders of magnitude smaller \cite{dr}, 
the elastic scattering between the (anti)neutrini and the photons may be safely omitted.

Ignoring the (presently unknown) (anti)neutrini masses, the interactions (a) of the (anti)neutrini of the electron generation of matter and (b) of the (anti)neutrini of the muon and of the $\tau$-lepton generations of matter 
are different: three of the graphs in Figs.~\ref{fig:NuNu}-\ref{fig:NuElPos} do not contribute to the collision rates of the muon and $\tau$-lepton (anti)neutrini. As the interactions of the muon and $\tau$-lepton (anti)neutrini 
are identical in the plasma, the study of the interactions induced only by $\nu_e$ and $\nu_\mu$ suffices for the purposes of this work.

In all cases, only Dirac neutrini are considered herein: therefore, each neutrino and its corresponding antineutrino are assumed to be distinguishable particles. The range of the available energy, corresponding to the temperatures 
$T_1$ and $T_2$, is such that no massive state may be produced `on shell', save for electrons and positrons. In addition, $-q^2$ does not exceed $300$ MeV$^2$ in this temperature range, a value which is at least seven orders 
of magnitude smaller than the square of the masses of the two IVBs. This property results in the simplification of the expressions for the various weak-interaction scattering amplitudes, and the subsequent dependence of the 
observables on only two physical constants, namely $G_F$ and $\xi \coloneqq \sin^2 \theta_W$, known as Fermi coupling constant and (square of the sine of the) weak-mixing angle, respectively. According to the most recent 
compilation of the physical constants by the Particle-Data Group \cite{pdg}, $G_F=1.1663787(6) \cdot 10^{-5}$ GeV$^{-2}$ and $\xi=0.23129(5)$. In this work, the differential cross sections (DCSs) and the total cross sections 
(TCSs) will be expressed as multiples of the representative weak-interaction cross section $\sigma_0 \coloneqq G^2_F s/\pi$. The $\sigma_0$ value, corresponding to $T=10^{11}$ K, is about $5 \cdot 10^{-42}$ cm$^2$ or $5 \cdot 10^{-18}$ b.

\section{\label{sec:Preliminary}The weak-interaction scattering amplitude}

The literature on the weak interaction is vast. For the sake of example, a thorough introduction to the subject may be obtained from Refs.~\cite{ah,gk}. I generally follow the formalism of Ref.~\cite{ah} herein. In this section, 
I only summarise those elements which are of interest in the narrow scope of this paper.

Inspection of Figs.~\ref{fig:NuNu}-\ref{fig:NuElPos} reveals that, in order to derive the various weak-interaction scattering amplitudes, one needs to obtain the leptonic currents applicable in the cases of the NC (vertices 
involving $Z^0$) and of the CC (vertices involving $W^\pm$). The latter are purely $V-A$ in character (`purely' implies equal - and opposite in sign - contributions of the vector and axial-vector components, ensuring the 
left-handedness of the neutrino and the right-handedness of the antineutrino), given for the $l \to \nu_l$ transition by the expression
\begin{equation*}
\bra{u(\nu_l)} J_\mu \ket{u(l)} = \frac{g}{\sqrt{2}} \bar{u}(\nu_l) \gamma_\mu \frac{1-\gamma_5}{2} u(l) \, \, \, ,
\end{equation*}
where the constant $g$ determines the overall strength of the CC processes. Of course, $u(l)$ is the spinor associated with the lepton $l$ (where $l$ denotes an electron, a muon, or a $\tau$-lepton) and $u(\nu_l)$ is the 
spinor associated with the corresponding neutrino; in this work, only electrons and electron neutrini are of relevance. The dependence of the spinors on the spin and on the $4$-momentum of the fermions is assumed (but not 
explicitly given). The normalisation condition of the spinors, associated with positive-energy solutions, is $u^\dagger(l) u(l)=2 E$, $E$ being the particle's total energy (where $l$ now stands for the neutrini as well). 
Assuming this normalisation condition, one can prove that $\bar{u}(l) u(l)=2 m$, $m$ being the particle's rest mass and $\bar{u}(l) \coloneqq u^\dagger(l) \gamma_0$.

On the other hand, the leptonic NC follows the generic form
\begin{equation*}
\bra{u(l)} J_\mu \ket{u(l)} = \frac{g_N}{2} \bar{u}(l) \gamma_\mu (g_V - g_A \gamma_5) u(l) \, \, \, ,
\end{equation*}
where the constant $g_N$ determines the overall strength of the NC processes, and $g_V$ and $g_A$ are known as vector and axial-vector couplings, respectively. These couplings are different for the neutrini and for the other 
leptons: the neutrino NC remains purely $V-A$ in character, i.e.,
\begin{equation*}
g_V^\nu = 1/2 \, \, \, , g_A^\nu = 1/2 \, \, \, , 
\end{equation*}
whereas for all other leptons
\begin{equation*}
g_V^l = -1/2 + 2\xi \, \, \, , g_A^l = -1/2 \, \, \, .
\end{equation*}

According to the Standard Model of the Electroweak Interactions, developed in the 1960s by Sheldon Lee Glashow, Abdus Salam, and Steven Weinberg, the couplings $g$ and $g_N$ are related via the expression: $g=g_N \xi$. A similar 
relation holds for the masses of the IVBs: $M_W = M_Z \xi$. As a result, $g/M_W=g_N/M_Z$. The Fermi coupling constant, the sole regulator of the strength of the weak-interaction processes in Fermi's current-current framework 
(pointlike approximation), is related to the aforementioned quantities via the expression
\begin{equation*}
\frac{G_F}{\sqrt{2}} = \frac{g^2}{8 M_W^2} = \frac{g_N^2}{8 M_Z^2} \, \, \, .
\end{equation*}

The last ingredient, which is necessary in order to advance to the derivation of the various weak-interaction scattering amplitudes, is the propagator associated with the virtual state; it is of the form
\begin{equation*}
\Pi^{\mu \nu} (q) = i \frac{-g^{\mu \nu} + q^\mu q^\nu / M^2}{q^2 - M^2} \, \, \, ,
\end{equation*}
where $M$ stands for the mass of the exchanged IVB.

Let me now combine these elements and obtain the weak-interaction scattering amplitude of one simple process, e.g., of the one involving the scattering of two neutrini of different flavour: $\nu_a \nu_b \to \nu_a \nu_b$. Such 
a process, featuring an electron neutrino as projectile, is schematically shown in Fig.~\ref{fig:NuNu}(c). The current-propagator-current form of the scattering amplitude reads as
\begin{align*}
\mathscr{T} & \coloneqq - i \bra{u(\nu_a)} J_\mu \ket{u(\nu_a)} \Pi^{\mu \nu} \bra{u(\nu_b)} J_\nu \ket{u(\nu_b)}\\
&= -i \frac{g_N}{2} \bar{u}(\nu_a) \gamma_\mu \frac{1 - \gamma_5}{2} u(\nu_a) i \frac{-g^{\mu \nu} + q^\mu q^\nu / M_Z^2}{q^2 - M_Z^2} \frac{g_N}{2} \bar{u}(\nu_b) \gamma_\nu \frac{1 - \gamma_5}{2} u(\nu_b)\\
&= \frac{g^2_N}{16} \bar{u}(\nu_a) \gamma_\mu (1 - \gamma_5) u(\nu_a) \frac{-g^{\mu \nu} + q^\mu q^\nu / M_Z^2}{q^2 - M_Z^2} \bar{u}(\nu_b) \gamma_\nu (1 - \gamma_5) u(\nu_b) \, \, \, ,
\end{align*}
which, for $q^2 \ll M_Z^2$, results in
\begin{align} \label{eq:EQ100}
\mathscr{T} &= \frac{g^2_N}{16 M_Z^2} \bar{u}(\nu_a) \gamma^\mu (1 - \gamma_5) u(\nu_a) \bar{u}(\nu_b) \gamma_\mu (1 - \gamma_5) u(\nu_b)\nonumber\\
&= \frac{G_F}{2\sqrt{2}} \bar{u}(\nu_a) \gamma^\mu (1 - \gamma_5) u(\nu_a) \bar{u}(\nu_b) \gamma_\mu (1 - \gamma_5) u(\nu_b) \, \, \, .
\end{align}

The square of the scattering amplitude $\lvert \mathscr{T} \rvert^2$, summed over the final states and properly averaged (for unpolarised cross sections) over the spin orientations in the initial state, is the backbone of the 
evaluation of the DCS of a process, complemented by the flux of the incident beam and the permissible Lorentz invariant phase space.

For massless projectiles, the DCS is obtained via the expression
\begin{equation*}
\frac{d\sigma}{d\tau} = \frac{\lvert \mathscr{T} \rvert^2}{2(s-m_t^2)} \, \, \, ,
\end{equation*}
where $d\tau$ denotes the infinitesimal Lorentz invariant phase space and $m_t$ stands for the rest mass of the target, i.e., $0$ for neutrini (Sections \ref{sec:D1}-\ref{sec:D5} and \ref{sec:D8}), and $m_e$ for electrons and 
positrons (Sections \ref{sec:D6} and \ref{sec:D7}). The infinitesimal Lorentz invariant phase space may be obtained via the relation
\begin{equation*}
d\tau = \frac{p_f d\Omega}{(4 \pi)^2 \sqrt{s}} \, \, \, ,
\end{equation*}
where $p_f$ denotes the modulus of the CM $3$-momentum in the final state. (Regarding the details relating to the Lorentz invariant phase space, see Ref.~\cite{ah}.) In Sections \ref{sec:D1}-\ref{sec:D5}, $p_f=\sqrt{s}/2$ and
\begin{equation} \label{eq:EQ105}
\frac{d\sigma}{d\Omega} = \frac{\lvert \mathscr{T} \rvert^2}{(8 \pi)^2 s} \, \, \, .
\end{equation}
In Sections \ref{sec:D6} and \ref{sec:D7}, $p_f=(s-m_e^2)/(2\sqrt{s})$ and Eq.~(\ref{eq:EQ105}) also holds. In Section \ref{sec:D8}, $p_f=\sqrt{s/4-m_e^2}$ and
\begin{equation*}
\frac{d\sigma}{d\Omega} = \frac{\lvert \mathscr{T} \rvert^2}{(8 \pi)^2 s} \sqrt{1-\frac{4 m_e^2}{s}} \, \, \, .
\end{equation*}

I will finally list a number of helpful relations obeyed by the matrix operator Tr, which returns the trace of a square matrix, i.e., the sum of its diagonal elements. Put forth by Richard Feynman, the trace technique 
comprises a set of operations enabling the speedy derivation of the sum of the contributions to the scattering amplitude from all spin orientations pertaining to the scattering process in question.
\begin{align*}
{\rm Tr} [A+B] &= {\rm Tr} [A] + {\rm Tr} [B]\\
{\rm Tr} [AB] &= {\rm Tr} [BA]\\
{\rm Tr} [ABC] &= {\rm Tr} [CAB] = {\rm Tr} [BCA]\\
{\rm Tr} [g_{ab}] &= 4 g_{ab}\\
{\rm Tr} [\gamma_a \gamma_b] &= {\rm Tr} [2 g_{a b} - \gamma_b \gamma_a] = {\rm Tr} [2 g_{a b} - \gamma_a \gamma_b] \Rightarrow {\rm Tr} [\gamma_a \gamma_b] = 4 g_{a b}\\
{\rm Tr} [\gamma_a \gamma_b \gamma_c \gamma_d] &= 4 \left( g_{ab} g_{cd} - g_{ac} g_{bd} + g_{ad} g_{bc} \right)\\
{\rm Tr} [\gamma_5 \gamma_a \gamma_b \gamma_c \gamma_d] &= 4 i \epsilon_{abcd}\\
\end{align*}
$A$, $B$, and $C$ denote square matrices of the same dimension. The traces of products of odd numbers of $\gamma$ matrices, as well as those of odd numbers of $\gamma$ matrices with the $\gamma_5$ matrix, vanish.

\section{\label{sec:Derivation}Derivation of the various weak-interaction differential cross sections}

In this section, the scattering amplitudes, as well as the DCSs and TCSs derived thereof, of the various scattering processes of the neutrini with the (anti)neutrini, as well as with electrons and positrons, will be extracted. 
Sections \ref{sec:D1}-\ref{sec:D5} deal with the channels in which only (anti)neutrini appear in the initial and final states. Sections \ref{sec:D6} and \ref{sec:D7} deal with the scattering of neutrini off electrons and 
positrons, respectively. Finally, Section \ref{sec:D8} pertains to the neutrino-antineutrino annihilation to an $e^+ e^-$ pair.

\subsection{\label{sec:D1}Scattering of neutrini of different flavour}

The scattering amplitude $\mathscr{T}$ for the interaction of neutrini of different flavour was obtained in Section \ref{sec:Preliminary}, see Eq.~(\ref{eq:EQ100}). For an electron neutrino as projectile, the graph of Fig.~\ref{fig:NuNu}(c) 
is relevant. As the neutrini are solely left-handed, the averaging over the spin orientations in the initial state is not applicable (the average of one value is the value itself). The square of the scattering amplitude 
$\lvert \mathscr{T} \rvert^2$ is obtained after multiplying $\mathscr{T}$ with its complex conjugate. However, given that $\mathscr{T}$ is one (complex) element (i.e., not a matrix), this operation is equivalent to using the 
Hermitian of $\mathscr{T}$, namely $\mathscr{T}^\dagger$, in the product.
\begin{equation*}
\lvert \mathscr{T} \rvert^2 \coloneqq \mathscr{T} \mathscr{T}^\dagger = \frac{G_F^2}{8} \, {\rm Tr} [\fsl{k}^\prime \gamma^\mu (1-\gamma_5) \fsl{k} \gamma^\nu (1-\gamma_5)] \, {\rm Tr} [\fsl{p}^\prime \gamma_\mu (1-\gamma_5) \fsl{p} \gamma_\nu (1-\gamma_5)]
\end{equation*}
The application of the first trace operator yields the projectile-related tensor
\begin{equation} \label{eq:EQ110}
K^{\mu \nu} = 8 \left( k^{\prime \, \mu} k^\nu + k^{\prime \, \nu} k^\mu + \frac{q^2}{2} g^{\mu\nu} - i \epsilon^{\mu \nu a b} k_a k^\prime_b \right) \, \, \, .
\end{equation}
The target-related tensor, resulting from the application of the second trace operator, is of similar structure
\begin{equation*}
P_{\mu \nu} = 8 \left( p^\prime_\mu p_\nu + p^\prime_\nu p_\mu + \frac{q^2}{2} g_{\mu\nu} - i \epsilon_{\mu \nu c d} p^c p^{\prime \, d} \right) \, \, \, .
\end{equation*}

It can be easily shown that $q_\mu K^{\mu \nu} = q_\nu K^{\mu \nu} = 0$. This property enables the simplification of the target-related tensor, via the introduction of its effective form
\begin{equation*}
(P_{\mu \nu})_{\rm eff} = 8 \left( 2 p_\mu p_\nu + \frac{q^2}{2} g_{\mu\nu} - i \epsilon_{\mu \nu c d} p^c q^d \right) \, \, \, .
\end{equation*}

The contraction of the tensors $K^{\mu \nu}$ and $(P_{\mu \nu})_{\rm eff}$ finally yields the value of $64 s^2$. Necessary in the extraction of this result are the relations: $s \coloneqq (k+p)^2 = 2 k \cdot p$, $u \coloneqq (k-p^\prime)^2 = -2 k \cdot p^\prime$, 
and $t \coloneqq (k-k^\prime)^2 = -2k \cdot k^\prime$, which are the expressions of the Mandelstam variables for massless initial and final states. One additional relation is needed, namely
\begin{equation*}
\epsilon_{\mu \nu c d} \epsilon^{\mu \nu a b} = - 2 \left( \delta^a_c \delta^b_d - \delta^b_c \delta^a_d \right) \, \, \, ,
\end{equation*}
where $\delta$ denotes the standard Kronecker delta, equal to $1$ for identical indices and $0$ for different ones. Evidently, 
\begin{equation} \label{eq:EQ120}
\lvert \mathscr{T} \rvert^2 = 8 G_F^2 s^2 \, \, \, .
\end{equation}

Using this result, one obtains the DCS
\begin{equation} \label{eq:EQ130}
\frac{d\sigma}{d\Omega} = \frac{G^2_F s}{8 \pi^2} = \frac{\sigma_0}{8 \pi}
\end{equation}
and, integrating over the solid angle, the TCS
\begin{equation*}
\sigma_{\rm tot} = \frac{\sigma_0}{2} \, \, \, .
\end{equation*}

\subsection{\label{sec:D2}Scattering of a neutrino off an antineutrino of different flavour}

For an electron neutrino as projectile, the graph of Fig.~\ref{fig:NuNuBar}(f) is relevant. Feynman's interpretation of `the negative-energy particle solutions propagating backward in time' as `the positive-energy antiparticle 
solutions propagating forward in time' provides a speedy solution to the scattering of this section from the result of the previous one: the current for the incident/outgoing antineutrino of Fig.~\ref{fig:NuNuBar}(f) moving 
forward in time is identical to the one for an outgoing/incident neutrino moving backward in time. This implies that the result, obtained for the graph Fig.~\ref{fig:NuNu}(c) (i.e., the result of Section \ref{sec:D1}), may be 
used, along with the obvious transformations $p \to - p^\prime$ and $p^\prime \to -p$, and enable the nearly effortless extraction of the DCS of this section. Evidently, one simply needs to interchange the Mandelstam variables 
$s$ and $u$ in the $\lvert \mathscr{T} \rvert^2$ result of Section \ref{sec:D1}. According to Eq.~(\ref{eq:EQ120}), only the Mandelstam variable $s$ enters $\lvert \mathscr{T} \rvert^2$, hence the result for the process of 
this section is
\begin{equation*}
\lvert \mathscr{T} \rvert^2 = 8 G_F^2 u^2 \, \, \, ,
\end{equation*}
leading to the DCS in the form
\begin{equation*}
\frac{d\sigma}{d\Omega} = \frac{G^2_F u^2}{8 \pi^2 s} = \frac{\sigma_0 u^2}{8 \pi s^2} \, \, \, ,
\end{equation*}
which, integrated over the solid angle, results in
\begin{equation*}
\sigma_{\rm tot} = \frac{\sigma_0}{6} \, \, \, .
\end{equation*}

\subsection{\label{sec:D3}Scattering of neutrini of the same flavour}

The peculiarity of this case is that the particles in the final state are indistinguishable. For the scattering of electron neutrini, the graphs of Fig.~\ref{fig:NuNu}(a) and (b) are relevant.

Starting from Eq.~(\ref{eq:EQ100}), one may write the scattering amplitude as
\begin{align} \label{eq:EQ140}
\mathscr{T} = \frac{G_F}{2\sqrt{2}} &\big( \bar{u}(\nu_a) \gamma^\mu (1 - \gamma_5) u(\nu_a) \bar{u}(\nu_b) \gamma_\mu (1 - \gamma_5) u(\nu_b)\nonumber\\
&+ \bar{u}(\nu_b) \gamma^\mu (1 - \gamma_5) u(\nu_a) \bar{u}(\nu_a) \gamma_\mu (1 - \gamma_5) u(\nu_b) \big) \, \, \, .
\end{align}
Let me rewrite the second term within the brackets in the form 
\begin{equation*}
( \bar{\chi} \gamma^\mu (1 - \gamma_5) \psi ) \, ( \bar{\psi} \gamma_\mu (1 - \gamma_5) \chi ) \, \, \, ,
\end{equation*}
where $\chi=u(\nu_b)$ (hence $\bar{\chi}=\bar{u}(\nu_b)$) and $\psi=u(\nu_a)$ (hence $\bar{\psi}=\bar{u}(\nu_a)$). By invoking two of the Fierz identities, namely
\begin{align} \label{eq:EQ150}
(\bar{\chi}\gamma^\mu \psi)(\bar{\psi} \gamma_\mu \chi)&=(\bar{\chi}\chi)(\bar{\psi}\psi)-\frac{1}{2}(\bar{\chi}\gamma^\mu \chi)(\bar{\psi}\gamma_\mu \psi)\nonumber\\
&-\frac{1}{2}(\bar{\chi}\gamma^\mu \gamma_5 \chi)(\bar{\psi}\gamma_\mu \gamma_5 \psi)-(\bar{\chi}\gamma_5 \chi)(\bar{\psi}\gamma_5 \psi) \, \, \, ,\nonumber\\
(\bar{\chi}\gamma^\mu \gamma_5 \psi)(\bar{\psi} \gamma_\mu \gamma_5 \chi)&=-(\bar{\chi}\chi)(\bar{\psi}\psi)-\frac{1}{2}(\bar{\chi}\gamma^\mu \chi)(\bar{\psi}\gamma_\mu \psi)\nonumber\\
&-\frac{1}{2}(\bar{\chi}\gamma^\mu \gamma_5 \chi)(\bar{\psi}\gamma_\mu \gamma_5 \psi)+(\bar{\chi}\gamma_5 \chi)(\bar{\psi}\gamma_5 \psi) \, \, \, ,
\end{align}
one obtains the result
\begin{equation*}
(\bar{\chi}\gamma^\mu (1-\gamma_5) \psi)(\bar{\psi} \gamma_\mu (1-\gamma_5) \chi) = -(\bar{\chi}\gamma^\mu (1-\gamma_5) \chi)(\bar{\psi}\gamma_\mu (1-\gamma_5) \psi)
\end{equation*}
and, after reverting to the original spinors,
\begin{equation*}
(\bar{u}(\nu_b)\gamma^\mu (1-\gamma_5) u(\nu_a))(\bar{u}(\nu_a) \gamma_\mu (1-\gamma_5) u(\nu_b)) = -(\bar{u}(\nu_b)\gamma^\mu (1-\gamma_5) u(\nu_b))(\bar{u}(\nu_a)\gamma_\mu (1-\gamma_5) u(\nu_a)) \, \, \, .
\end{equation*}
Of course, the wavefunction of fermions is antisymmetric under particle exchange, hence
\begin{equation*}
-(\bar{u}(\nu_b)\gamma^\mu (1-\gamma_5) u(\nu_b))(\bar{u}(\nu_a)\gamma_\mu (1-\gamma_5) u(\nu_a)) = + (\bar{u}(\nu_a)\gamma^\mu (1-\gamma_5) u(\nu_a)) (\bar{u}(\nu_b)\gamma_\mu (1-\gamma_5) u(\nu_b)) \, \, \, .
\end{equation*}

Inserting this relation into Eq.~(\ref{eq:EQ140}), one obtains
\begin{align*}
\mathscr{T} = \frac{G_F}{2\sqrt{2}} &\big( \bar{u}(\nu_a) \gamma^\mu (1 - \gamma_5) u(\nu_a) \bar{u}(\nu_b) \gamma_\mu (1 - \gamma_5) u(\nu_b)\\
&+ \bar{u}(\nu_a)\gamma^\mu (1-\gamma_5) u(\nu_a) \bar{u}(\nu_b)\gamma_\mu (1-\gamma_5) u(\nu_b) \big)\\
= \frac{G_F}{\sqrt{2}} &\bar{u}(\nu_a) \gamma^\mu (1 - \gamma_5) u(\nu_a) \bar{u}(\nu_b) \gamma_\mu (1 - \gamma_5) u(\nu_b) \, \, \, .
\end{align*}
Therefore, the amplitude for the scattering of neutrini of the same flavour is twice the result of Eq.~(\ref{eq:EQ100}), i.e., twice the scattering amplitude for the scattering of neutrini of different flavour. Consequently, 
there is no need to repeat the calculation of Section \ref{sec:D1} in order to obtain the result of this section; all one needs to do is multiply the DCS of Eq.~(\ref{eq:EQ130}) by $4$. Finally,
\begin{equation} \label{eq:EQ160}
\frac{d\sigma}{d\Omega} = \frac{4 G^2_F s}{8 \pi^2} = \frac{\sigma_0}{2 \pi} \, \, \, .
\end{equation}

Attention is needed when deriving the TCS for the scattering of neutrini of the same flavour from the result of Eq.~(\ref{eq:EQ160}). Owing to the fact that the final state comprises indistinguishable particles, the integration 
of the DCS between the $\theta$ limits of $0$ and $\pi$ yields an erroneous result! The outgoing particles are indistinguishable and it is fallacious to attach labels to them~\footnote{I am indebted to Carlo Guinti for drawing 
my attention to this subtle point and for bringing forth the argument on the integration limits of $\theta$.}, e.g., by identifying particle (1) as the one scattered at angle $\theta$ and particle (2) as the one scattered at 
$\pi-\theta$. The indistinguishability of the particles in the final state requires that the integration be performed over the solid angle of $2 \pi$, i.e., over a hemisphere, thus yielding the TCS result
\begin{equation*}
\sigma_{\rm tot} = \sigma_0 \, \, \, .
\end{equation*}

To be able to compare the DCS of this section with those obtained for the other neutrino-induced processes, one may proceed in either of two ways:
\begin{itemize}
\item by making use of the DCS of Eq.~(\ref{eq:EQ160}), also bearing in mind that the corresponding TCS must involve an integration over a hemisphere,
\item by halving the DCS of Eq.~(\ref{eq:EQ160}) and integrating over $4 \pi$.
\end{itemize}
I will follow the latter option. Of course, all expressions, obtained from the $\lvert \mathscr{T} \rvert^2$ result of this section (e.g., on the basis of interchanges of $4$-momenta) for other processes, a) must involve the 
original $\lvert \mathscr{T} \rvert^2$ result and b) must correspond to TCSs involving the integration of the corresponding DCSs over $4 \pi$. The restricted solid-angle domain pertains exclusively to processes yielding 
indistinguishable particles in the final state, which (for Dirac neutrini) is the case only in this section.

\subsection{\label{sec:D4}Elastic scattering of a neutrino off an antineutrino of the same flavour}

For the elastic scattering of an electron neutrino off an electron antineutrino, the graphs of Fig.~\ref{fig:NuNuBar}(a) and (c) are relevant; the latter graph represents the annihilation to a neutrino-antineutrino pair of 
the same flavour. The substitutions $p \to - p^\prime$ and $p^\prime \to -p$ enable the speedy extraction of the DCS of this section from the $\lvert \mathscr{T} \rvert^2$ result of Section \ref{sec:D3}. Evidently,
\begin{equation} \label{eq:EQ180}
\frac{d\sigma}{d\Omega} = \frac{G^2_F u^2}{2 \pi^2 s} = \frac{\sigma_0 u^2}{2 \pi s^2} \, \, \, ,
\end{equation}
which, integrated over the solid angle, results in
\begin{equation*}
\sigma_{\rm tot} = \frac{2 \sigma_0}{3} \, \, \, .
\end{equation*}

\subsection{\label{sec:D5}Neutrino-antineutrino annihilation to a neutrino-antineutrino pair of different flavour}

For an electron neutrino-antineutrino pair, the graph of Fig.~\ref{fig:NuNuBar}(d) is relevant. Straightforward considerations, again invoking Feynman's interpretation of the negative-energy particle solutions, lead to a result 
identical to the one obtained in Section \ref{sec:D2}. The DCS is of the form
\begin{equation*}
\frac{d\sigma}{d\Omega} = \frac{\sigma_0 u^2}{8 \pi s^2} \, \, \, ,
\end{equation*}
which, integrated over the solid angle, results in
\begin{equation*}
\sigma_{\rm tot} = \frac{\sigma_0}{6} \, \, \, .
\end{equation*}

\subsection{\label{sec:D6}Scattering of neutrini off an electron}

For the scattering of an electron neutrino off an electron, the graphs of Fig.~\ref{fig:NuElPos}(a) and (b) are relevant. The scattering amplitude is of the form
\begin{align} \label{eq:EQ200}
\mathscr{T} &= \left( \frac{g_N}{2 M_Z} \right)^2 \bar{u}(\nu_e) \gamma^\mu \frac{1 - \gamma_5}{2} u(\nu_e) \bar{u}(e^-) \gamma_\mu (g_V^l - g_A^l \gamma_5) u(e^-)\nonumber\\
&+ \left( \frac{g}{\sqrt{2} M_W} \right)^2 \bar{u}(\nu_e) \gamma^\mu \frac{1 - \gamma_5}{2} u(e^-) \bar{u}(e^-) \gamma_\mu \frac{1 - \gamma_5}{2} u(\nu_e)\nonumber\\
&= \frac{G_F}{\sqrt{2}} \big( \bar{u}(\nu_e) \gamma^\mu (1 - \gamma_5) u(\nu_e) \bar{u}(e^-) \gamma_\mu (g_V^l - g_A^l \gamma_5) u(e^-)\nonumber\\
&+ \bar{u}(e^-) \gamma^\mu (1 - \gamma_5) u(\nu_e) \bar{u}(\nu_e) \gamma_\mu (1 - \gamma_5) u(e^-) \big) \, \, \, .
\end{align}
After employing the two Fierz identities of Eqs.~(\ref{eq:EQ150}), the second term within the brackets may be rewritten as $\bar{u}(\nu_e) \gamma^\mu (1 - \gamma_5) u(\nu_e) \bar{u}(e^-) \gamma_\mu (1 - \gamma_5) u(e^-)$, 
resulting in
\begin{align} \label{eq:EQ220}
\mathscr{T} &= \frac{G_F}{\sqrt{2}} \bar{u}(\nu_e) \gamma^\mu (1 - \gamma_5) u(\nu_e) \bar{u}(e^-) \gamma_\mu \left( g_V^l + 1 - (g_A^l +1) \gamma_5 \right) u(e^-)\nonumber\\
&= \frac{G_F}{\sqrt{2}} \bar{u}(\nu_e) \gamma^\mu (1 - \gamma_5) u(\nu_e) \bar{u}(e^-) \gamma_\mu (C_V - C_A \gamma_5) u(e^-) \, \, \, .
\end{align}
where $C_V \coloneqq g_V^l + 1$ and $C_A \coloneqq g_A^l +1$ for the scattering of an electron neutrino off an electron. Equation (\ref{eq:EQ220}) provides an explanation for the differences in the interactions between electron 
and muon/$\tau$-lepton neutrini with the electron. As, below the temperature $T_1$, the CM energy is not sufficient for the creation of muons and $\tau$-leptons in the final state, the CC graph contributes to the scattering 
amplitude only in case of an incident electron neutrino. To be able to apply the results of the calculation also in the case of incident muon and $\tau$-lepton neutrini, I will retain the constants $C_V$ and $C_A$ (applicable 
for the scattering of an electron neutrino off an electron) and bear in mind that, in the case of an incident $\nu_{\mu,\tau}$, they must be replaced by $\widetilde{C}_V = C_V-1 = g_V^l$ and $\widetilde{C}_A = C_A-1 = g_A^l$. Evidently,
\begin{align*}
\lvert \mathscr{T} \rvert^2 = &\frac{G_F^2}{2} \, {\rm Tr} [\fsl{k}^\prime \gamma^\mu (1-\gamma_5) \fsl{k} \gamma^\nu (1-\gamma_5)]\\
&\frac{1}{2} \, {\rm Tr} [(\fsl{p}^\prime+m_e) \gamma_\mu (C_V-C_A \gamma_5) (\fsl{p}+m_e) \gamma_\nu (C_V-C_A\gamma_5)] \, \, \,
\end{align*}
where the factor $1/2$ in front of the second trace takes account of the averaging of the spin orientations in the initial state (target electron). The application of the first trace operator yields the projectile-related 
tensor of Eq.~(\ref{eq:EQ110}). The application of the second trace yields the target-related tensor, which (owing to the fact that the electron NC is not purely $V-A$ in character) is now of a more complex form.
\begin{equation*}
P_{\mu \nu} = 4 \left( c_1 \left( p^\prime_\mu p_\nu + p^\prime_\nu p_\mu + \left( \frac{q^2}{2} - m_e^2 \right) g_{\mu\nu} \right) - i c_2 \epsilon_{\mu \nu c d} p^c p^{\prime \, d} + c_3 m_e^2 g_{\mu\nu} \right) \, \, \, ,
\end{equation*}
where $c_1=C_V^2+C_A^2$, $c_2=2 C_V C_A$, and $c_3=C_V^2-C_A^2$.

As in Section \ref{sec:D1}, an effective target-related tensor may be constructed on the basis of the properties: $q_\mu K^{\mu \nu} = q_\nu K^{\mu \nu} = 0$.
\begin{equation*}
(P_{\mu \nu})_{\rm eff} = 4 \left( c_1 \left( 2 p_\mu p_\nu + \left( \frac{q^2}{2} - m_e^2 \right) g_{\mu\nu} \right) - i c_2 \epsilon_{\mu \nu c d} p^c q^d + c_3 m_e^2 g_{\mu\nu} \right) \, \, \, .
\end{equation*}
The contraction of $K^{\mu\nu}$ of Eq.~(\ref{eq:EQ110}) and $(P_{\mu \nu})_{\rm eff}$ of the previous equation results in
\begin{equation} \label{eq:EQ240}
\lvert \mathscr{T} \rvert^2 = 4 G_F^2 \left( C_1 (s-m^2_e)^2 + C_2 (u - m^2_e)^2 + 2 C_3 m^2_e t \right) \, \, \, ,
\end{equation}
where $C_1=c_1+c_2=(C_V+C_A)^2$, $C_2=c_1-c_2=(C_V-C_A)^2$, and $C_3=c_3=(C^2_V-C^2_A)$. As aforementioned, this result is valid for an electron neutrino as projectile. The same formula holds for muon and $\tau$-lepton 
neutrini as projectiles, but the constants must be redefined, following the replacements $C_V \to \widetilde{C}_V$ and $C_A \to \widetilde{C}_A$. (The constant $C_2$ is not affected by these substitutions.)

The DCS for the scattering of neutrini off an electron is of the form
\begin{equation} \label{eq:EQ225}
\frac{d\sigma}{d\Omega} = \frac{\sigma_0}{16 \pi s^2} \left( C_1 (s-m^2_e)^2 + C_2 (u - m^2_e)^2 + 2 C_3 m^2_e t \right) \, \, \, ,
\end{equation}
where (as already explained) the constants $C_1$ and $C_3$ depend on the neutrino flavour. After integrating the DCS over the solid angle, one obtains
\begin{equation} \label{eq:EQ230}
\sigma_{\rm tot}=\frac{\sigma_0}{4} \left( 1 - \frac{m^2_e}{s} \right)^2 \left( C_1 + C_2 \frac{s^2 + m^4_e + s m^2_e}{3 s^2} - C_3 \frac{m^2_e}{s} \right) \, \, \, ,
\end{equation}
which is applicable for $s \geq m_e^2$. These expressions are in agreement with those derived in the pioneering paper of Herrera and Hacyan \cite{herr}.

\subsection{\label{sec:D7}Scattering of neutrini off a positron}

For the scattering of an electron neutrino off a positron, the graphs of Fig.~\ref{fig:NuElPos}(c) and (d) are relevant. Following the line of argumentation of the previous section, the graph of Fig.~\ref{fig:NuElPos}(d) does 
not contribute to the scattering amplitude in case of incident muon and $\tau$-lepton neutrini. The scattering amplitude may be obtained from Eq.~(\ref{eq:EQ240}) after the interchange $s \leftrightarrow u$, which is equivalent 
to the interchange $C_1 \leftrightarrow C_2$ in Eqs.~(\ref{eq:EQ225}) and (\ref{eq:EQ230}). Therefore, the DCS for the scattering of neutrini off a positron reads as
\begin{equation*}
\frac{d\sigma}{d\Omega} = \frac{\sigma_0}{16 \pi s^2} \left( C_1 (u - m^2_e)^2 + C_2 (s-m^2_e)^2 + 2 C_3 m^2_e t \right) \, \, \, ,
\end{equation*}
where the constants $C_1$ and $C_3$ depend on the neutrino flavour as explained in the previous section. After integrating the DCS over the solid angle, one obtains
\begin{equation*}
\sigma_{\rm tot}=\frac{\sigma_0}{4} \left( 1 - \frac{m^2_e}{s} \right)^2 \left( C_1 \frac{s^2 + m^4_e + s m^2_e}{3 s^2} + C_2 - C_3 \frac{m^2_e}{s} \right) \, \, \, ,
\end{equation*}
which is applicable for $s \geq m_e^2$. As expected, these expressions are in agreement with those given in Ref.~\cite{herr}.

\subsection{\label{sec:D8}Neutrino-antineutrino annihilation to an $e^+ e^-$ pair}

For the annihilation of an electron neutrino-antineutrino pair to an $e^+ e^-$ pair, the graphs of Fig.~\ref{fig:NuNuBar}(b) and (e) are relevant. The scattering amplitude, corresponding to the neutrino-antineutrino annihilation 
to an $e^+ e^-$ pair, may be obtained from the $\lvert \mathscr{T} \rvert^2$ result of Section \ref{sec:D6} after the replacements $k^\prime \to -p$, $p \to -p^\prime$, and $p^\prime \to k^\prime$. These replacements are 
equivalent to the substitutions $s \to u$, $u \to t$, and $t \to s$. One additional issue requires attention. Equation (\ref{eq:EQ240}) for $\lvert \mathscr{T} \rvert^2$ was obtained after averaging over the spin orientations 
of the target electron; no averaging is to be performed in the case of the neutrino-antineutrino annihilation, hence the amplitude, obtained after the aforementioned substitutions are performed, must be multiplied by $2$. The 
DCS thus becomes
\begin{equation*}
\frac{d\sigma}{d\Omega} = \frac{\sigma_0}{8 \pi s^2} \sqrt{1-\frac{4 m^2_e}{s}} \left( C_1 (u - m^2_e)^2 + C_2 (t - m^2_e)^2 + 2 C_3 m^2_e s \right) \, \, \, ,
\end{equation*}
where the constants have been defined in Section \ref{sec:D6}. The integration of the DCS over the solid angle yields
\begin{equation*}
\sigma_{\rm tot}=\sigma_0 \sqrt{1-\frac{4 m^2_e}{s}} \left( \frac{C_1+C_2}{6} \left( 1 - \frac{m^2_e}{s} \right) + C_3 \frac{m^2_e}{s} \right) \, \, \, ,
\end{equation*}
which is applicable for $s \geq 4 m_e^2$. As Kuznetsov and Savin \cite{ks} noticed (and proposed this property as a simple criterion for judging the correctness of relevant calculations~\footnote{Of course, the fulfilment of 
this condition may be a necessary, but is not a sufficient condition for the correctness of the calculations.}), the TCS for the neutrino-antineutrino annihilation to an $e^+ e^-$ pair does not depend on the constant $C_A$ in 
the vicinity of $s=4m_e^2$; it comes out equal to $C_V^2 \sqrt{\epsilon} \sigma_0 / 2$, where $s = 4 m_e^2 (1 + \epsilon)$ and $\epsilon$ is positive and small (compared to $1$).

\section{\label{sec:Summary}Summary of the results}

\begin{figure}
\begin{center}
\includegraphics[width=15cm]{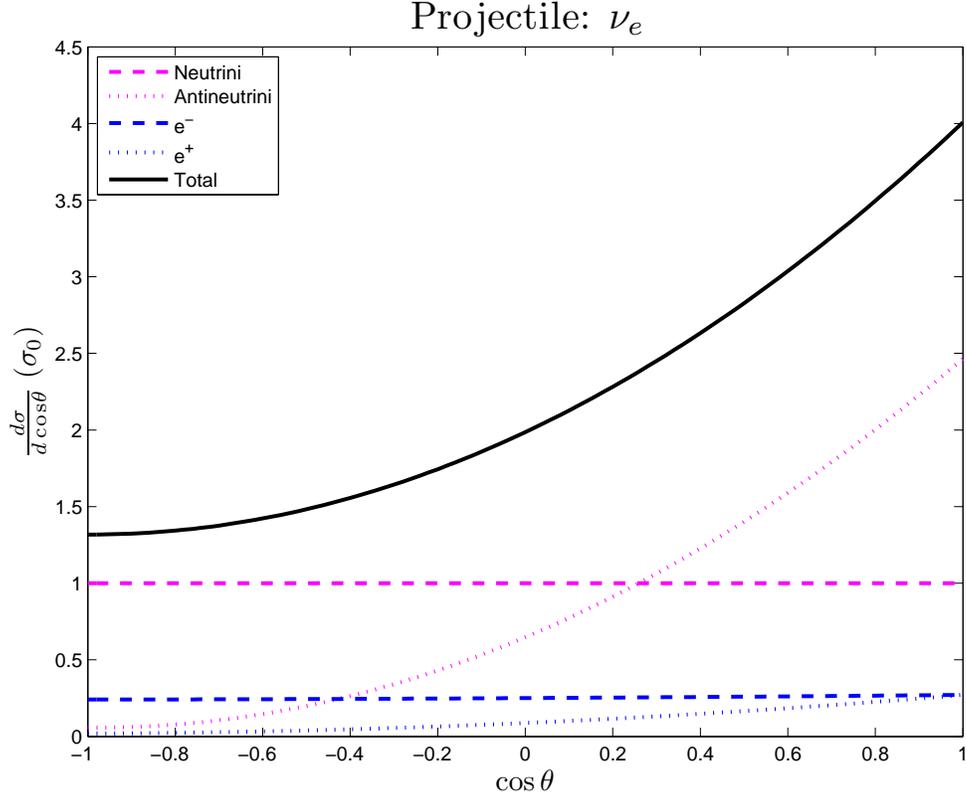}
\caption{\label{fig:ElectronNeutrino}Plot of the angular distributions of the cross section, detailed in Sections \ref{sec:D1}-\ref{sec:D8}, for an electron neutrino as projectile. The quantity $\sigma_0$ has been defined at 
the end of Section \ref{sec:FD}. The differential cross section for antineutrino targets does not vanish at $\cos\theta=-1$ because it includes the annihilation channel to an $e^+ e^-$ pair. The square of the CM energy for 
the incident particles corresponds to a temperature of $3 \cdot 10^{10}$ K.}
\vspace{0.35cm}
\end{center}
\end{figure}

\begin{figure}
\begin{center}
\includegraphics[width=15cm]{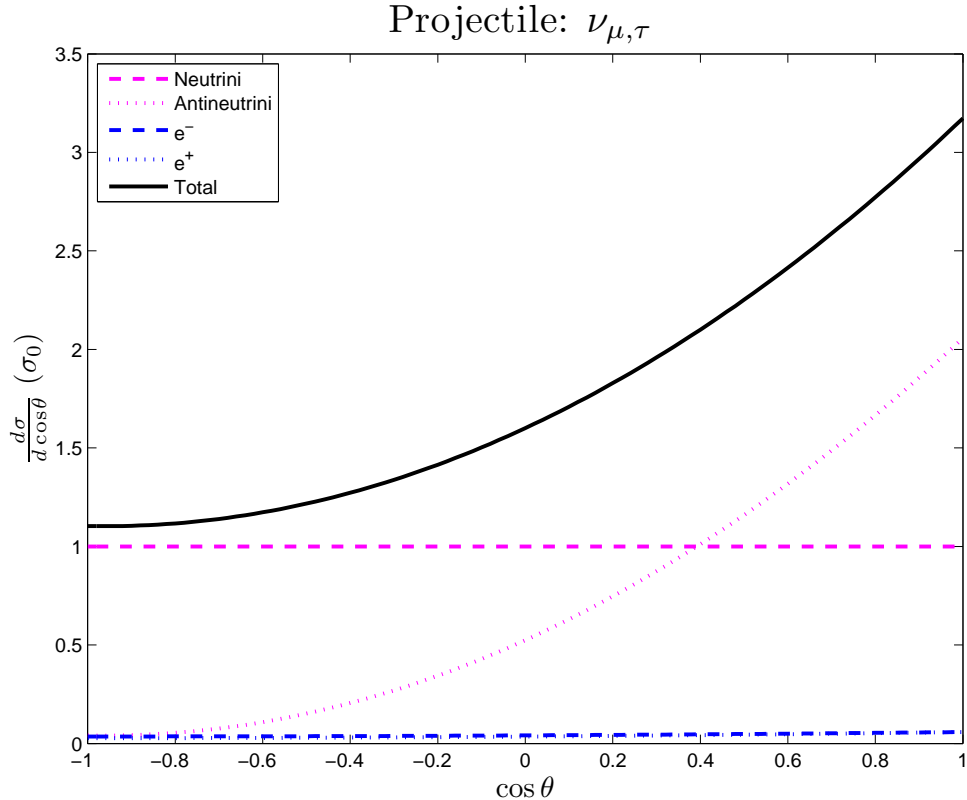}
\caption{\label{fig:MuonTauNeutrino}Plot of the angular distributions of the cross section, detailed in Sections \ref{sec:D1}-\ref{sec:D8}, for muon/$\tau$-lepton neutrini as projectiles. The quantity $\sigma_0$ has been 
defined at the end of Section \ref{sec:FD}. The differential cross section for antineutrino targets does not vanish at $\cos\theta=-1$ because it includes the annihilation channel to an $e^+ e^-$ pair. The square of the CM 
energy for the incident particles corresponds to a temperature of $3 \cdot 10^{10}$ K.}
\vspace{0.35cm}
\end{center}
\end{figure}

Figures \ref{fig:ElectronNeutrino} and \ref{fig:MuonTauNeutrino} provide plots of the angular distributions of the cross section, detailed in Sections \ref{sec:D1}-\ref{sec:D8}, separately for electron and muon/$\tau$-lepton 
neutrini as projectiles. Visual inspection of these figures leaves no doubt that the interactions of the (anti)neutrini amongst themselves are more important than those involving electron and positron targets~\footnote{Due to 
the spin degeneracy of the electrons and the positrons, the contributions of the graphs involving electron and positron targets to the mean-free path of the neutrini in simulations of the early Universe are twice as large as 
one obtains from Figs.~\ref{fig:ElectronNeutrino} and \ref{fig:MuonTauNeutrino} (as well as from Table \ref{tab:TCSs}); due to the two possible spin orientations in case of electrons and positrons, twice as many of these 
particles may be `packed together' in a given volume, at a given temperature, as (anti)neutrini of each of the three flavours.}.

All the TCSs, derived in Section \ref{sec:Derivation}, are listed in Table \ref{tab:TCSs}. In the three last cases (corresponding to the results obtained in Sections \ref{sec:D6}-\ref{sec:D8}), expressions derived under the 
high-energy approximation ($s \gg m^2_e$) are quoted in the table.

\begin{table}
{\bf \caption{\label{tab:TCSs}}}Total cross sections (TCSs) for the interactions of the (anti)neutrini with the various ingredients of the plasma in the early Universe. The TCSs are expressed as multiples of the representative 
weak-interaction cross section $\sigma_0$, which has been defined at the end of Section \ref{sec:FD}. In the three last cases, the formulae are high-energy approximations; the exact expressions are given in Section \ref{sec:Derivation}. 
The TCS for the neutrino-antineutrino annihilation ($\nu_a \bar{\nu}_a$) contains all the final-state channels, save for the annihilation to an $e^+ e^-$ pair, which is shown separately in last row of the table.
\vspace{0.2cm}
\begin{center}
\begin{tabular}{|c|c|c|c|}
\hline
Target $\downarrow$, Projectile $\to$ & $\nu_e$ & $\nu_\mu$ & $\nu_\tau$\\
\hline
$\nu_e$ & $1$ & $1/2$ & $1/2$\\
$\nu_\mu$ & $1/2$ & $1$ & $1/2$\\
$\nu_\tau$ & $1/2$ & $1/2$ & $1$\\
$\bar{\nu}_e$ & $1$ & $1/6$ & $1/6$\\
$\bar{\nu}_\mu$ & $1/6$ & $1$ & $1/6$\\
$\bar{\nu}_\tau$ & $1/6$ & $1/6$ & $1$\\
$e^-$ & $\frac{1}{4}+\xi+\frac{4 \xi^2}{3}$ & $\frac{1}{4}-\xi+\frac{4 \xi^2}{3}$ & $\frac{1}{4}-\xi+\frac{4 \xi^2}{3}$\\
$e^+$ & $\frac{1}{3} \left( \frac{1}{4}+\xi+4 \xi^2 \right)$ & $\frac{1}{3} \left( \frac{1}{4}-\xi+4 \xi^2 \right)$ & $\frac{1}{3} \left( \frac{1}{4}-\xi+4 \xi^2 \right)$\\
\hline
$e^+ e^-$ creation & $\frac{1}{6} \left( 1+4 \xi+8 \xi^2 \right)$ & $\frac{1}{6} \left( 1-4 \xi+8 \xi^2 \right)$ & $\frac{1}{6} \left( 1-4 \xi+8 \xi^2 \right)$\\
\hline
\end{tabular}
\end{center}
\vspace{0.5cm}
\end{table}

\section{\label{sec:Discrepancies}Discrepancies in the literature}

Comparing the expressions of this paper with those appearing in a number of published works (peer-reviewed articles and books) reveals a number of discrepancies. I will next address in chronological order those of the 
discrepancies which I am aware of.
\begin{itemize}
\item Flowers and Sutherland \cite{fs} calculated the $\nu_e \nu_e$ DCS and TCS correctly, including the appropriate integration of the DCS for indistinguishable particles in the final state. However, their expression for the 
$\nu_e \bar{\nu}_e$ DCS, see their Eq.~(8), is fourfold the value obtained from Eq.~(\ref{eq:EQ180}) of this work.
\item Hannestad and Madsen \cite{hm} presented in tabular form the scattering amplitudes corresponding to the various neutrino-induced processes relevant to the decoupling. The entry for the scattering of neutrini of the same 
flavour in their tables yields a DCS which is half the result of this work. The same problem appeared in Tables 5.1 and 5.2 of Ref.~\cite{hh}, as well as in Table 1 of Ref.~\cite{gg}. Dolgov also commented on these mismatches 
in Ref.~\cite{dolg} (p.~356).
\item Xing and Zhou \cite{xz} gave the expressions for the $\nu_a \nu_b$ and $\bar{\nu}_a \bar{\nu}_b$ TCSs for $a=b$ and $a \neq b$ (pp.~39-40). Those expressions are in agreement with the results of Table~\ref{tab:TCSs}. They 
then advanced to treat the TCS for the $\nu_a \bar{\nu}_a$ processes and came up with a value which is twice as large as the $\nu_a \nu_a$ TCS. Unfortunately, they obtained that result after employing the wrong DCS; evidently, 
they did not replace the Mandelstam variable $s$ with $u$ in $\lvert \mathscr{T} \rvert^2$ when extracting the $\nu_a \bar{\nu}_a$ scattering amplitude from the one they had obtained for the $\nu_a \nu_a$ process.
\item Lesgourgues, Mangano, Miele, and Pastor \cite{lmmp} presented in tabular form the TCSs corresponding to the various neutrino-induced processes relevant to the decoupling (see their Tables 1.7 and 1.8). To the best of my 
knowledge, this is the only place in the literature where an observable (e.g., the TCS) is presented in tabular form for all the processes studied in this work. Nevertheless, their tables are particularly worrying. To start 
with, all the TCSs off (anti)neutrino targets are half the corresponding entries of Table \ref{tab:TCSs} of this work. In addition, their TCSs for the neutrino-antineutrino annihilation to an $e^+ e^-$ pair are twice the 
corresponding entries of Table \ref{tab:TCSs} of this work. (Interestingly, their TCSs for the scattering of neutrini off electrons or positrons agree with the results of this work!) These discrepancies are puzzling.
\end{itemize}

Additional discrepancies (relating to the DCSs and TCSs for the neutrino-antineutrino annihilation to an $e^+ e^-$ pair) were reported by Kuznetsov and Savin in Ref.~\cite{ks}. As my overview in the domain of Neutrino Physics 
is limited, the chances are that the aforementioned list is anything but exhaustive. I would be indebted to the colleagues who could communicate further discrepancies to me; I will acknowledge such contributions in future 
versions of this paper.

\section{\label{sec:Conclusions}Conclusions}

The present study dealt with the neutrino-induced processes relevant to the physics of the early Universe, namely with the interactions of the neutrini and the antineutrini of the three generations of matter amongst themselves, 
as well as with the electrons and the positrons of the plasma. These processes are of interest in other domains too, namely in the physics of compact stars. All the differential cross sections of these processes were derived, 
hopefully in a didactical and elucidating manner, following the standard methodology of Ref.~\cite{ah}. To facilitate the overview and the crosscheck of the results, the corresponding total cross sections are shown in tabular 
form (Table \ref{tab:TCSs}). The angular distributions of the sums of the DCSs, detailed in Sections \ref{sec:D1}-\ref{sec:D8}, are shown in Fig.~\ref{fig:ElectronNeutrino} and \ref{fig:MuonTauNeutrino}, separately for electron 
and muon/$\tau$-lepton neutrini as projectiles.

Discrepancies were reported with some scientific publications and books comprising part of the literature on this subject. By no means should the list of discrepancies, addressed in this work, be considered as exhaustive.

The present study deals only with Dirac neutrini; its extension to also cover Majorana neutrini may be worth pursuing.

\begin{ack}
I am grateful to Alexander V.~Kuznetsov for having clarified a question regarding the results of Ref.~\cite{herr} and to Joseph A.~Formaggio for drawing my attention to Ref.~\cite{gk}. It would be an unpardonable omission to 
fail acknowledging the exchange of constructive electronic mail with Carlo Giunti on a number of interesting subjects relating to this paper.

All Feynman graphs in this work were created with the software package JaxoDraw \cite{jd}, available from http://jaxodraw.sourceforge.net/.
\end{ack}


\begin{thebibliography}{99}
\bibitem{dr} D.~A.~Dicus, W.~W.~Repko, `Photon-neutrino interactions', Phys.~Rev.~Lett.~79 (1997) 569--571.
\bibitem{pdg} C.~Patrignani \etal~(Particle Data Group), `The review of Particle Physics', Chin.~Phys.~C 40 (2016) 100001.
\bibitem{ah} I.~J.~Aitchison, A.~J.~Hey, `Gauge Theories in Particle Physics', Vol.~2, 3rd edn., IoP Publishing, 2004.
\bibitem{gk} C.~Giunti, C.~W.~Kim, `Fundamentals of Neutrino Physics and Astrophysics', Oxford University Press, 2007.
\bibitem{herr} M.~A.~Herrera, S.~Hacyan, `Relaxation time of neutrinos in the early Universe', Astrophys.~J.~336 (1989) 539--543.
\bibitem{ks} A.~V.~Kuznetsov, V.~N.~Savin, `The cross-section of the neutrino-antineutrino pair conversion into the electron-positron pair and its use in astrophysical calculations: correction of mistakes', Conference on 
`Physics of Fundamental Interactions', Moscow, MEPhI, November 17-21, 2014; available online from http://www.icssnp.mephi.ru/content/file/2014/section8/8\_16\_Kuznetsov\_MEPhI\_14e.pdf.
\bibitem{fs} E.~G.~Flowers, P.~G.~Sutherland, `Neutrino-neutrino scattering and supernovae', Astrophys.~J.~208 (1976) L19--L21.
\bibitem{hm} S.~Hannestad, J.~Madsen, `Neutrino decoupling in the early Universe', Phys.~Rev.~D 52 (1995) 1764--1769.
\bibitem{hh} S.~Hannestad, `Aspects of Neutrino Physics in the Early Universe', Ph.D.~dissertation, University of Aarhus, 1997.
\bibitem{gg} N.~Y.~Gnedin, O.~Y.~Gnedin, `Cosmological neutrino background revisited', Astrophys.~J.~509 (1998) 11--15.
\bibitem{dolg} A.~D.~Dolgov, `Neutrinos in Cosmology', Phys.~Rep.~370 (2002) 333--535.
\bibitem{xz} Zhizhong Xing, Shun Zhou, `Neutrinos in Particle Physics, Astronomy and Cosmology', Springer, 2011.
\bibitem{lmmp} J.~Lesgourgues, G.~Mangano, G.~Miele, S.~Pastor, `Neutrino Cosmology', Cambridge University Press, 2013.
\bibitem{jd} D.~Binosi, L.~Theu\ss{}l, `JaxoDraw: A graphical user interface for drawing Feynman diagrams', Comput.~Phys.~Commun.~161 (2004) 76--86.
\end{thebibliography}
\end{document}